\documentclass[11pt]{article}
\usepackage{amssymb}
\advance \topmargin by -\headheight
\advance \topmargin by -\headsep
\evensidemargin \oddsidemargin
\catcode`\@=11
\def\marginnote#1{}
\def\numberbysection{\@addtoreset{equation}{section}
        \def\theequation{\thesection.\arabic{equation}}}
\def\underline#1{\relax\ifmmode\@@underline#1\else
        $\@@underline{\hbox{#1}}$\relax\fi}
\def\titlepage{\@restonecolfalse\if@twocolumn\@restonecoltrue\onecolumn
     \else \newpage \fi \thispagestyle{empty}\c@page\z@
        \def\thefootnote{\fnsymbol{footnote}} }
\def\endtitlepage{\if@restonecol\twocolumn \else  \fi
        \def\thefootnote{\arabic{footnote}}
        \setcounter{footnote}{0}}  
\numberbysection

\begin{document}
\title{Baxter T-Q Equation for Shape Invariant Potentials.
The Finite-Gap Potentials Case}
\author{Ovidiu Lipan \thanks{email: olipan@hsph.harvard.edu} \\
     Harvard University, HSPH,\\
     655 Huntington Avenue,\\
     Boston, MA 02115, USA \\
    \and
    Constantin Rasinariu \thanks{email: crasinariu@popmail.colum.edu}\\
    Columbia College Chicago,\\
    600 South Michigan Ave,\\
    Chicago, IL 60605, USA}
\maketitle
\begin{abstract}
The Darboux transformation applied recurrently on a Schroedinger operator
generates what is called a {\em dressing chain}, or from a different point
of view, a set of supersymmetric shape invariant potentials. The finite-gap
potential theory is a special case of the chain. For the finite-gap case,
the equations of the chain can be expressed as a time evolution of a
Hamiltonian system. We apply Sklyanin's method of separation of variables to
the chain. We show that the classical equation of the separation of
variables is the Baxter T-Q relation after quantization.
\end{abstract}

{\bf Key Words:} Baxter Q-operator, Shape Invariant Potentials

{\bf PACS:} 71.23.Ft; 03.65.F; 03.65.Bz; 02.20.-a

\parskip 5pt
\section{Introduction}

What is the most universal method of solving completely integrable models? From
Sklyanin's point of view \cite{Sklyanin1} it is the separation of variables in its
most general form. Therefore, it is desirable to understand old methods of
integration in the light of the modern approach to the separation of variables. One
of the most important ``old'' techniques applied with success to many physical and
mathematical problems is the Darboux transformation. In short, the two main themes
of this paper can be best described by two keywords: the Darboux transformation and
the Sklyanin method of separation of variables. The Darboux transformation is, for
example, a fundamental tool in the supersymmetric approach to quantum mechanics and
in the theory of dressing chains \cite{Veselov,Cooper}. Though the model is
different from the DST model studied in \cite{Sky} this paper and \cite{Sky} are
closely related. The Sklyanin method aims to connect the separation of variables as
we know it from the Hamiltonian mechanics with the new techniques of exactly solving
mathematical physics problem, namely the Inverse Scattering Method and its quantum
version \cite{Kuz}. The worked example in this paper is the dressing chain
representation of the finite-gap potential theory. We review the main ideas from the
Darboux transformations with an emphasis on the Hamiltonian view-point on the
finite-gap theory, following A.P. Veselov and A.B. Shabat \cite{Veselov}. We then
work the Sklyanin method \cite{Sklyanin1,Sklyanin2,Sklyanin3} on the finite-gap
theory. We find the canonical separated variables for the dressing chain
representation of the finite-gap potential theory. From the Quantum Inverse
Scattering Theory point of view, the equation for the separated variables is the
classical version of the Baxter T-Q equation. Then the quantum version of the
finite-gap theory is presented, together with the corresponding R-matrix and the
Baxter Q-operator. The conclusions and outlook will close the paper.

 \section{The Darboux transformation and the Hamiltonian approach for the
 finite-gap theory}

Consider the Schroedinger operator for a potential $u(x)$, ${\cal H }=-D^2+u(x)$
where $D\equiv d/dx$. Factorize it as a product of two first order operators
\begin{equation}
\label{Sr}
{\cal H}=A^{*}A ~,
\end{equation}
where $A=D-f(x)$ and $A^{*}=-D-f(x).$
The Darboux transformation sends ${\cal H}_1=A^{*}A$ into ${\cal H}_2=AA^{*}+\alpha \mathbf{1}$
where $\mathbf{1}$ is the identity operator and $\alpha $ is a constant. The functions $f_1$ and
$f_2$ obtained from the factorization
\begin{equation}
\label{fac}
{\cal H}_i=-(D+f_i)(D-f_i) ~,
\end{equation}
are related by the equation
\begin{equation}
\label{eq_f}
(f_1+f_2)^{'}=f_1^2-f_2^2+\alpha ~,
\end{equation}
where the prime means differentiation with respect to $x$.
From the supersymmetric quantum mechanics point of view, equation
(\ref{eq_f}) is exactly the shape invariance condition \cite{Cooper},
written in terms of the superpotentials $W_i=-f_i$ .

We can continue this process of factorization and Darboux transformation and
obtain a chain of equations
\begin{equation}
\label{chain}
(f_i+f_{i+1})^{'}=f_i^2-f_{i+1}^2+\alpha_i\;\;\;i=1,2,...\;.
\end{equation}
This chain is called a {\it dressing chain}. In what follows we will
consider only periodic chains
\begin{equation}
\label{per}
f_i=f_{i+N},\;\;\;\alpha_i=\alpha _{i+N} ~,
\end{equation}
where the period $N$ is a positive integer. Supersymmetrically speaking,
periodic chains correspond to the cyclic shape invariant potentials
\cite{Sukhatme}. The properties of the dressing chain depend drastically on
the period $N$ and the sum $\alpha=\alpha_1+ \dots +\alpha_N.$ There are
four cases to be considered depending if $N$ is even or odd and $\alpha $ is
equal to zero or not. The finite-gap theory, to be studied in this paper,
corresponds to the case $N$ odd and $\alpha=0.$ In this case the chain is a
completely integrable Hamiltonian system. We will use for the case
$\alpha = 0$ instead of the constant $\alpha_i$, the constants $\beta_i$
related with $\alpha$ by $\beta_i - \beta_{i+1} = \alpha_i$.
 If we regard the variable $x$ in
$f_i(x)$ as a {\it time} variable, then the dressing chain expresses the
time evolution of the variables $f_i(x),\;\;i=1 \dots N$. Their evolution is
generated by the Hamiltonian
\begin{equation}
\label{H_evol}
H=\sum_{i=1}{N}\left(\frac{1}{3}f_i^3+\beta_if_i\right) ~,
\end{equation}
with the Poisson bracket
\begin{eqnarray}
\label{Poisson_f}
\{f_i,f_j\}&=&(-1)^{(j-i)mod N} ~,\\
\{f_i,f_i\}&=&0~.
\end{eqnarray}
This Poisson bracket is not canonical. To obtain canonical variables, first
let us denote by $g_i$ a set of variables defined by
\begin{equation}
\label{g}
g_i=f_i+f_{i+1}
\end{equation}
with the Poisson structure
\begin{equation}
\label{Poisson_g}
\{g_i,g_{i-1}\}=1 ~.
\end{equation}
all other brackets being zero.
Though the variables $g_i$ seem to be redundant at this point, later on they
will prove to be useful. Now, the canonical variables $(X_i,x_i)$
\begin{equation}
\label{Pisson_x}
\{X_i,X_j\}=\{x_i,x_j\}=0\;,\;\; \{X_i,x_j\}=\delta_{ij} ~,
\end{equation}
will generate the Poisson structure for the variables $g_i$ if
\begin{equation}
\label{g_x}
g_i=X_i+x_{i+1}~.
\end{equation}
As we mentioned before, the finite-gap case ($N=2n+1=$ odd) is completely
integrable. Therefore the Hamiltonian responsible for the time evolution
belongs to a set $H_0,H_1,...,H_n$ of independent involutive Hamiltonians.
To show this, we use the method of inverse scattering theory. The Lax matrix
build on canonical variables $(X_i,x_i)$ is
\begin{equation}
\label{Lax1}
 L_i^{(x)}(u)=\left(\begin{array}{cc}
          x_i      & 1  \\
           x_iX_i+\beta_i+u      &  X_i
         \end{array}\right )\;\;,
\end{equation}
where $(x) \equiv (X_i,x_i)$ and $u$ is a complex parameter. Then we
construct the monodromy matrix
\begin{equation}
\label{monoL}
L^{(x)}(u)=\prod_{i=N}^{1} L^{(x)}_i(u)
                   =L^{(x)}_N(u) L^{(x)}_{N-1}(u) \ldots L^{(x)}_1(u)
\end{equation}
and take its trace:
\begin{equation}
\label{monod}
\tau _N(u)=Tr L(u).
\end{equation}
The trace $\tau_N(u)$ generates the set of involutive Hamiltonians
\begin{equation}
\label{Ham_gen}
\tau _N(u)=H_1u^n+H_3u^{n-1}+ \dots +H_{2N+1} ~.
\end{equation}
The Hamiltonian $H_3$ is just the Hamiltonian for the chain. The fact that
the set of the Hamiltonians is involutive, is a consequence of the classical
r-matrix identity. Denoting by $id_2$ the unit $2\times 2$ matrix and
introducing the notations for the tensor products $l^{(1)}=l\otimes id_2$,
$l^{(2)}=id_2\otimes l$, we have
\begin{equation}
\label{r-matrix}
 \{l_i^{(1)}(u_1),l_i^{(2)}(u_2)\}
               =[r_{12}(u_1-u_2),l_i^{(1)}(u_1)l_i^{(2)}(u_2)]\delta_{ij}~,
\end{equation}
where
\begin{equation}
\label{r}
r_{12}(u)=-\frac{1}{u}{\cal P}_{12}~,
\end{equation}
and ${\cal P}_{12}$ is the permutation operator in ${\bf C}^2\otimes {\bf
C}^2.$ It is important to notice that although the Lax matrix is written in
terms of the variable $(X,x)$, the Hamiltonians $H_i$ (\ref{Ham_gen}) depend
only on the variables $g_i$, $i=1,2,\ldots,N$. Moreover, it is possible to
generate the Hamiltonians $H_i$ as a trace of a monodromy matrix written
directly in terms of the chain variables $f_i$
\begin{equation}
\label{mono_f}
F(u)=\prod_{i=1}^N \left(\begin{array}{cc}
          f_i      & 1  \\
           f_i^2+\beta_i+u      &  f_i
         \end{array}\right )\;\;,
\end{equation}
\begin{equation}
\label{tau_f}
\tau _N=Tr F(u)\;.
\end{equation}
The connection of the dressing chain with the finite-gap theory for the
Schroedinger operators is described in \cite{Veselov}. For the finite-gap
theory see also \cite{Novikov1,Novikov2} and the references therein.  Here
we will emphasize only those notions which will be important later on. To
each solution $f_i(x)$ of the dressing chain, or in other words, to each
solution of the time evolution of the Hamiltonian system, corresponds a
sequence of $N$ finite-gap Schroedinger operators
\begin{equation}
\label{Sch_fin}
{\cal H}_i=-D^2+u_i(x) ~,
\end{equation}
where the potentials $u_i(x)$ are given by
\begin{equation}
\label{pot}
u_i=f_i^{'}+f_i^2\;.
\end{equation}
The spectral curve of these operators is the spectral curve of the monodromy
matrix $F(u)$
\begin{equation}
\label{spec_curve}
det(F(\lambda)-\mu \, \mathbf{1})=0 ~,
\end{equation}
where $ \mathbf{1}$ is the unit matrix, and it can be written as
\begin{equation}
\label{spec2}
\mu^2-\tau _N(\lambda) \, \mu-\prod _{i=1}^N(\lambda +\beta_i)=0 ~.
\end{equation}
{}From the Darboux transformation we can obtain a recurrence relation
between the logarithmic derivatives of the Bloch eigenfunctions. The Bloch
eigenfunctions $\psi _i$ of the operator ${\cal H}_i$ are given by
\begin{equation}
\label{Bloch}
{\cal H}_i\psi _i=(\lambda+\beta _i)\psi _i ~.
\end{equation}
Due to the Darboux transformation, two successive Bloch eigenfunctions are
connected through
\begin{equation}
\label{con}
\psi _{i+1}=(D-f_i)\psi _i ~.
\end{equation}
This implies that for the logarithmic derivatives $\chi _i=D\ln \psi_i$ we
have the recurrence
\begin{equation}
\label{rec_chi}
\chi_i=f_i+\frac{\beta_i+\lambda}{f_i+\chi_{i+1}} ~.
\end{equation}

\section{Sklyanin method of separation of variables}
To understand the Sklyanin method \cite{Sklyanin1,Sklyanin2,Sklyanin3,Kuz},
let us start with an old example. For an Hamiltonian of the form
\begin{equation}
\label{oldHam}
H=\frac{1}{2}(p_1^2+..+p_n^2)+\frac{1}{2}(\omega_1q_1^2+...+\omega_nq_n^2)\;,
\end{equation}
we notice that the variables are not only {\it canonical},
$\{p_i,q_j\}=\delta_ {ij}$ but also {\it separated} i.e. each pair
$(p_i(t),q_i(t))$ lies on the curve
\begin{equation}
\label{curve}
p_i^2+\omega_iq_i^2=const ~.
\end{equation}
In general, let us consider a Hamiltonian system having $d$ degrees of
freedom and integrable in Liouville's sense. This means that it is given a
$2d$-dimensional symplectic manifold and $d$ independent Hamiltonians $H_i$
in involution
\begin{equation}
\label{H_inv}
\{H_i,H_j\}=0,\;\;\;i,j=1 \ldots d\;.
\end{equation}
A system of canonical variables $\lambda\equiv \{\lambda_i\}_{i=1}^d$ and
$\mu\equiv \{\mu_i\}_{i=1}^d$ satisfying
\begin{equation}
\label{Candeg}
\{\lambda_i,\lambda_j\}=\{\mu_i,\mu_j\}=0~, \quad \{\lambda_i,\mu_j\}=\delta_{ij}
\end{equation}
will be called {\it separated} if there exists $d$ relations of the form
\begin{equation}
\label{sep}
W_j(\lambda_j,\mu_j,H_1,...,H_d)=0 ~.
\end{equation}
For the dressing chain the variables $(X,x)$ are canonical but not
separated. We can raise then the question of how to find {\it canonical
separated} variables for the dressing chain. If the integrable system is
solvable via the Inverse Scattering Method then we can use a method proposed
by Sklyanin to find the transformation from the canonical variables $(X,x)$
to the canonical separated variables $(\lambda,\mu)$. The desired
transformation will be obtained as a composition of B\"acklund
transformations. The next section is devoted to the B\"acklund
transformation for the dressing chain.

\section{B\"acklund transformations }
Following Sklyanin, we need to find a canonical transformation (we will use
also the name B\"acklund transformation) from the variables $(X,x)$ to
$(Y,y)$. The important property is that the canonical transformation will
depend on the spectral parameter $\lambda $. This parameter $\lambda $ will
allow us to find at the end the canonical separated variables. Being a
canonical transformation, the Poisson structure and the set of Hamiltonians
$H_i$ must remain unchanged when expressed in the variable $(Y,y)$. Since
Lax matrix $L(u)$ is a monodromy matrix
\begin{equation}
\label{Mo}
L^{(x)}(u)=L_N^{(x)}(u) \cdots L_2^{(x)}(u)L_1^{(x)}(u) ~,
\end{equation}
we can transform the Lax matrices at each site $L_i^{(x)}(u)$
\begin{equation}
\label{tra}
M_i(u-\lambda)L_i^{(x)}(u)=L_i^{(y)}(u)M_{i-1}(u-\lambda) ~,
\end{equation}
because the trace $\tau _N(u)$ is invariant due to
$M_N(u-\lambda)L^{(x)}(u)=L^{(y)}(u)M_N(u-\lambda)$. To keep the same
Poisson structure we ask that the matrices $M_i(u)$ should obey the same
Poisson bracket (\ref{r-matrix}) as $L_i(u)$ obeys. Practically, we first
have to choose one out of many matrices which obeys the r-matrix Poisson
bracket (\ref{r-matrix}) and then be lucky enough to find that (\ref{tra})
has a solution $Y(X,x), y(X,x)$ for every spectral parameter $u$. The
solution will depend on the parameter $\lambda $ which is exactly what we
want. The technical way to solve (\ref{tra}) is quite interesting. The idea
is to use, besides the phase spaces $(X,x)$ and $(Y,y)$, two more spaces:
$(S,s)$ and $(T,t)$. The phase spaces $(S,s)$ and $(T,t)$ are auxiliary
spaces, the B\"acklund transformation being between $(X,x)$ and $(Y,y)$.
With the help of these auxiliary spaces, we can write a version of
(\ref{tra}) as
\begin{equation}
\label{trast}
M_i^{(s)}(u-\lambda)L_i^{(x)}(u)=L_i^{(y)}(u)M_i^{(t)}(u-\lambda)~.
\end{equation}
To go back to (\ref{tra}) we simply need to impose the constraints
\begin{equation}
\label{cons}
t_i=s_{i-1}\;\; , \;\;T_i=S_{i-1} ~.
\end{equation}
We apply now the above method to the dressing chain. The Lax matrix is
\begin{equation}
\label{Lax11}
 L_i^{(x)}(u)=\left(\begin{array}{cc}
          x_i      & 1  \\
           x_iX_i+\beta_i+\lambda      &  X_i
         \end{array}\right )\;\;.
\end{equation}
Here $(x)$ stands for the pair of variables $(X,x)$ and $u$ is the spectral
parameter. We choose the matrix $M$ to be identical with $L$. Because in
(\ref{trast}) the index $i$ is the same on both sides, we can drop it and
write the matrix equation as
\begin{equation}
\label{mateq}
\left(\begin{array}{cc}
          s      & 1  \\
           u-\lambda +sS    &  S
         \end{array}\right )\left(\begin{array}{cc}
          x      & 1  \\
           u+\beta+xX     &  X
         \end{array}\right )=\left(\begin{array}{cc}
          y     & 1  \\
           u+\beta+yY    &  Y
         \end{array}\right )\left(\begin{array}{cc}
          t      & 1  \\
           u-\lambda+tT    &  T
         \end{array}\right )\;\;.
\end{equation}
The solution to the system is
\begin{eqnarray}
\label{solution2}
S&=&-x+\xi ~,\\
X&=&-s+\frac{\lambda +\beta}{t-x} ~,\\
T&=&-y+\frac{\lambda +\beta}{t-x} ~,\\
Y&=&-t+\xi ~.
\end{eqnarray}
We remark that $\xi$ is a free variable. This is a consequence of the fact
that the conserved Hamiltonians depend only on the combination
$X_i+x_{i+1}$. The generating function $F_{\lambda}(yt,xs)$ is
\begin{equation}
\label{gener}
F_{\lambda}(yt,xs)=y(t-\xi)-(x-\xi)s-(\lambda+\beta)\ln(t-x) ~,
\end{equation}
from which we get
\begin{eqnarray}
\label{deriv}
X=\frac{\partial F_\lambda}{\partial x}\;&,&\;
S=\frac{\partial F_\lambda}{\partial s}~,\\
Y=-\frac{\partial F_\lambda}{\partial y}\;&,&\;
T=-\frac{\partial F_\lambda}{\partial t} ~.
\end{eqnarray}
To simplify the formulas we choose the $\xi=x$. The generating function
becomes
\begin{equation}
\label{Fs}
F_{\lambda}=y(t-x)-(\lambda +\beta)\ln(t-x) ~.
\end{equation}
The constrains $t_i=s_{i-1}\;,\;T_i=S_{i-1}$  give:
\begin{equation}
\label{si1}
X_i=-s_i+\frac{\lambda +\beta_i}{s_{i-1}-x_i}\\
\end{equation}
which can be solved, in principle, for $s_i$. Then
\begin{equation}
\label{si2}
Y_i+y_{i+1}=X_{i+1}+x_i+s_{i+1}-s_{i-1} ~.
\end{equation}
In terms of the old variables $g_i=X_i+x_{i+1}$ the transformation reads:
\begin{equation}
\label{Bac}
    g_i=z_i-\frac{\lambda +\beta _{i}}{z_{i-1}} ~,
\end{equation}
\begin{equation}
\label{Bac-tilde}
    {\tilde g}_i=g_{i+1}+z_{i-1}-z_{i+1} ~,
\end{equation}
where ${\tilde g}_i$ are the transformed variables and
\begin{equation}
\label{z}
 z_i=x_{i+1}-s_i ~.
\end{equation} Note that (\ref{Bac}) and (\ref{Bac-tilde}) are just the
canonical transformations we were looking for. To obtain the
concrete form of these transformations we need to solve
(\ref{Bac}) for $z_i$, and then to use these values to find
${\tilde g}_i$. Because for an arbitrary $\lambda$ equation
(\ref{Bac}) cannot be explicitly solved, we leave the canonical
transformation in an implicit form. However for special values of
$\lambda$, the canonical transformation can be explicitly solved,
as we are going to exemplify in the section 10, eqs. (10.2),
(10.3) and (10.4).

The first equation in (\ref{Bac}) is the discrete Riccati equation. It can
be linearized with the help of the following change of variables:
\begin{equation}
\label{linear}
z_i=\frac{\psi _{i+1}}{\psi_i}\;.
\end{equation}
We obtain
\begin{equation}
\label{Ric}
\psi_{i+1}=\psi_ig_i+(\lambda+\beta_i)\psi_{i-1}\;\;,~~ i=0, \ldots ,N ~.
\end{equation}
The periodic boundary condition $z_0=z_{N}$ implies
\begin{equation}
\label{cond_psi}
\psi _1\psi_{N-1}=\psi_0\psi_N\;.
\end{equation}
{}From (\ref{Fs}), the generating function for the canonical transformation
is
\begin{eqnarray}
\label{gen_can2}
X_i&=&\frac{\partial \Phi_\lambda}{\partial x_i} ~,\\
Y_i&=&-\frac{\partial \Phi_\lambda}{\partial y_i} ~,
\end{eqnarray}
\begin{equation}
\label{gen_can1}
\Phi_\lambda({\vec y},{\vec x})=\sum_{i=1}^N F_\lambda(y_is_{i-1}|x_is_i)
       =\sum_{i=1}^Ny_i(s_{i-1}-x_i)-(\lambda +\beta_i)\ln (s_{i-1}-x_i)~.
\end{equation}
Here we denote by ${\vec x}=(x_1, \ldots , x_N)$. Finally in terms of $z_i$
the generating function can be written as
\begin{equation}
\label{gen_z}
\Phi_\lambda({\vec y},{\vec x})
         =\sum_{i=1}^Ny_i(-z_{i-1})-(\lambda +\beta_i)\ln (-z_{i-1})~.
\end{equation}

\section{Canonical transformations and Darboux factorization}
To decipher the meaning of the variables $z_i$ present in the canonical
transformation (\ref{Bac}) we will use the knowledge obtained from the
Darboux method of factorization. First, from (\ref{Bac}), find $z_{i-1}$
\begin{equation}
-z_{i-1}=\frac{\lambda +\beta _i}{g_i-z_i} ~.
\end{equation}
Then, compare this result with formula (\ref{rec_chi}) that gives the
recurrence relation between two logarithmic derivatives of the Bloch
eigenfunctions
\begin{equation}
\chi _i=f_i+\frac{\lambda +\beta_i}{f_i+\chi _{i+1}} ~.
\end{equation}
We obtain thus:
\begin{equation}
\label{chi}
z_{i-1}=f_i-\chi_i ~,
\end{equation}
which in terms of the superpotentials $W_i = -f_i$ reads as
$$
-z_i = W_i + \chi_i ~.
$$
Therefore the variable $z_i$ taken with a minus sign, is the sum between the
superpotential $W_i$ and the logarithmic derivative of the Bloch
eigenfunction. It is interesting to obtain the time evolution of the
variables $z_i$. From
\begin{eqnarray}
\label{timez}
-\psi_i^{''}+u_i\psi_i&=&(\lambda+\beta_i)\psi_i ~,\\
u_i&=&f_i^{'}+f_i^2 ~,\\
z_{i-1}^{'}&=&f_i^{'}-(\psi_i^{'}/\psi_i)^{'} ~
\end{eqnarray}
 we get
\begin{equation}
\label{evolz}
z_{i-1}^{'}=\chi_i^2-f_i^2+\lambda+\beta_i ~.
\end{equation}

\section{Separated canonical variables}
At this point we have the canonical B\"acklund transformations (\ref{Bac}).
Let us use the symbol $B_{\lambda}$ for this transformation. Our goal is to
find separated canonical variables. Following Sklyanin, we consider the
composition $B_{\lambda_1...\lambda_N}=B_{\lambda_1}\circ...\circ
B_{\lambda_N}$ of B\"acklund transformations and the corresponding
generating function $F_{\lambda _1...\lambda _N}(y,x).$ If we treat $\lambda
$'s as dynamical variables and $y$'s as parameters then
$F_{\lambda_1...\lambda_N}(y,x)$ becomes the generating function of the
$N$-parametric canonical transformation from $(X,x)$ to $(\mu,\lambda)$
given by
\begin{equation}
\label{cans}
X_i=\frac{\partial F_{\lambda_1...\lambda_N}}{\partial x_i} ~,\quad
\mu_i=-\frac{\partial F_{\lambda_1...\lambda_N}}{\partial \lambda_i} ~.
\end{equation}
This transformation is not only canonical but also separates the variables.
See \cite{Kuz} for details.

Each pair $(\lambda_i,\mu_i)$ lie on a curve given implicitly by
\begin{equation}
\label{cu}
W(\lambda_i,\mu_i)=0\;.
\end{equation}
To find the curve W we use the parameter $\mu $ which is the variable
conjugated to $\lambda $
\begin{equation}
\label{mu}
\mu =-\frac{\partial F_{\lambda}}{\partial \lambda}\;,
\end{equation}
and search for a function $f(\mu )$ such that
\begin{equation}
\label{fmu}
det(f(\mu)-L(\lambda))=0 ~.
\end{equation}
Then the spectral curve W is
\begin{equation}
W(\lambda_i,f(\mu_i);{H_i})\equiv det(f(\mu_i)-L(\lambda_i))=0 ~.
\end{equation}

For the dressing chain
\begin{equation}
\label{f}
f(\mu)=-e^{\mu} ~.
\end{equation}
To prove this, we will show that $-e^{\mu}$ is an {\it eigenvalue} for
$L(\lambda)$ so the property $det(f(\mu)-L(\lambda))=0$ is immediate.

{}From the definition of $\mu$ (\ref{mu}) and from the generating function
(\ref{gen_z}) we get
\begin{equation}
\label{zpr}
z_1\cdots z_n=-e^{\mu} ~.
\end{equation}
Now, by a  simple computation
\begin{equation}
\label{eigenL}
L^{(x)}_i(\lambda) \left(\begin{array}{c}
          1   \\
           -s_{i-1}
         \end{array}\right )=z_{i-1}\left(\begin{array}{c}
          1   \\
           -s_{i}
         \end{array}\right )\;\;,
\end{equation}
This proves that for $L=L_N\cdots L_1$ the eigenvalue is $z_1\cdots z_N.$

The spectral curve $W(\lambda,\mu)$ can be expressed in terms of the trace
$\tau_N$ and $\beta_i$
\begin{equation}
\label{spec}
det(v-L(\lambda))=v^2-\tau_N(\lambda)\,v+\prod_{i=1}^N(\lambda+b_i)
\end{equation}
so
\begin{equation}
\label{Wdres}
e^{2\mu}+\tau_N(\lambda)\,e^{\mu}-\prod_{i=1}^N(\lambda+b_i)=0 ~.
\end{equation}
Here
\begin{equation}
\label{taun}
\tau _N(\lambda)=H_1\lambda^n+H_3\lambda^{n-1}+ \ldots + H_{2N+1}~,
\end{equation}
where $N=2n+1$ and $H_1, H_3,\ldots,H_{2N+1}$ are integrals of the chain.

\section{Quantum case}

To get the quantum version of the theory described so far we will use the
R-matrix approach. This will ensure the commutativity of the Hamiltonians
$H_i$ after quantization. {}From classical variables $(x,X)$ we move to the
quantum variables $(x,\partial_x).$ The local quantum Lax matrix
\begin{equation}
\label{quntL}
L(u|x,\partial_x)=\left(\begin{array}{cc}
          x      & 1  \\
           u+x\partial_x     &  \partial_x
         \end{array}\right )
\end{equation}
verifies the quantum commutation relation
\begin{equation}
\label{YB}
R_{12}(u_1-u_2)L^{(1)}(u_1)L^{(2)}(u_2)=L^{(2)}(u_2)L^{(1)}(u_1)R_{12}(u_1-u_2) ~,
\end{equation}
where
\begin{equation}
\label{RR}
R_{12}(u)=u+{\cal P}_{12}
\end{equation}
is the $SL(2)$-invariant solution to the quantum Yang-Baxter equation
\cite{Faddeev}. The monodromy operator and its trace are defined like in the
classical case. The commutativity of the Hamiltonians $H_i$
\begin{equation}
\label{comH}
[H_i,H_j]=0 ~,
\end{equation}
is a consequence of (\ref{YB}). The whole machinery of the Quantum Inverse
Scattering Method can be put to work at this stage. We will limit to study
only the Baxter Q-operator and the Baxter T-Q relation. The Q-operator will
depend upon the spectral parameter $\lambda .$ Let us denote it by
$Q(\lambda ).$ The interesting aspect is that the classical B\"acklund
transformation $B_\lambda $ is the classical limit of the similarity
transformation
\begin{equation}
\label{O}
{\cal O}\to Q(\lambda ){\cal O} Q^{-1}(\lambda ) ~.
\end{equation}
For details see \cite{Kuz}. In the next section we will explicitly
construct $Q(\lambda)$ as an integral operator.

\section{$Q$-operator}

For the Baxter $Q$-operator we require the three usual properties. First, it
has to commute with the trace of the monodromy matrix $\tau_N(u) =
\prod_{i=N}^1 L(u|x_i,\partial_{x_i}) $
\begin{equation}
\label{tauQ}
[\tau_N(u),Q(\lambda )]=0 ~,
\end{equation}
second, it has to commute with itself
\begin{equation}
\label{QQ}
[Q(\lambda_1),Q(\lambda_2)]=0 ~,
\end{equation}
and the last important property imposed is the Baxter T-Q equation, i.e. the
$Q$-operator should satisfy a finite difference equation
\begin{equation}
\label{TQ}
\tau_N(\lambda )Q(\lambda )=A(\lambda)Q(\lambda -1)+B(\lambda)Q(\lambda +1)
\end{equation}
where $A(\lambda )$ and $B(\lambda )$ are two functions (not operators) of
the spectral parameter $\lambda$ .

We will follow \cite {Sky} and construct $Q(\lambda)$ as an integral
operator
\begin{equation}
\label{Q_i}
(Q(\lambda )\psi)({\vec x})
       =\int d{\vec t}\int d{\vec y}
       \prod _{i=1}^N R_{\lambda +\beta_i-1}(t_i,x_i|t_{i-1}y_i)\psi({\vec y}) ~.
\end{equation}
Here $d{\vec t}=dt_N...dt_1$ and similar for alike symbols. If we introduce
the ${\cal R}$-operator as
\begin{equation}
\label{intR}
({\cal R}_\lambda \psi )(s,x)=\int dy \int dt R_{\lambda}(s,x|t,y)\psi (y) ~,
\end{equation}
the formula (\ref{Q_i}) can be understand in the general sense of the trace
of a monodromy matrix
\begin{equation}
\label{Qtr}
Q(\lambda)=Tr_{t_N}{\cal R}_{\lambda +\beta_N-1}^1...{\cal R}_{\lambda +\beta_1-1}^N ~.
\end{equation}

In the notation $R_{\lambda}(t,y|s,x)$ we recognize the ${\it auxiliary }$
indexes $s,t$ and the ${\it quantum}$ indexes $x,y$. The $Q$-operator can be
expressed as an integral operator
\begin{equation}
\label{intQ}
(Q(\lambda ) \psi)({\vec x})
     =\int dy_1...\int dy_N {\cal Q}_\lambda ({\vec x}|{\vec y})\psi ({\vec y})
\end{equation}
with the kernel
\begin{equation}
\label{K}
{\cal Q}_\lambda ({\vec x}|{\vec y})=\int dt_N...\int dt_1 \prod_{i=N}^1
    R_{\lambda +\beta_i-1}(t_{i-1},x_i|t_i,y_i) ~.
\end{equation}
After this general introduction, we move forward to find the concrete form
of the operator ${\cal R}_\lambda$. The first property of the Baxter
Q-operator, namely the commutation $[\tau_N(u),Q(\lambda )]=0$ is fulfilled
if ${\cal R}_\lambda $ is a solution of an equation similar to (\ref{YB})
\begin{equation}
\label{MYB}
M(u-\lambda |s,\partial_s) L(u|x,\partial_x){\cal R}_\lambda={\cal R}_\lambda
L(u|y,\partial_y)M(u-\lambda |t,\partial_t) ~,
\end{equation}
where $L(u|x,\partial_x)$ is the local quantum Lax matrix (\ref{quntL}) and
$M(u-\lambda)$ is another matrix which obeys the quantum commutation
(\ref{YB}.) The main difficulty is how to chose the matrix $M(u-\lambda)$ so
that, the equation (\ref{MYB}) for ${\cal R}(\lambda)$ has a solution for
every complex parameter $u$ and by the other hand the $Q$-operator thus
obtained has the required properties. The second property of the
$Q$-operator comes from the Yang-Baxter equation which can be obtained from
(\ref{MYB}) by a standard technique, see \cite{Faddeev}. Returning to
equation (\ref{MYB}) we take $M$ to be of the same form as the Lax matrix
(\ref{quntL}). We obtain
\begin{eqnarray}
\label{mlquatum1}
\left(\begin{array}{cc}
           s      & 1  \\
           u-\lambda +s\partial_s    &  \partial_s
         \end{array}\right )\left(\begin{array}{cc}
          x      & 1  \\
           u+x\partial_x     &  \partial_x
         \end{array}\right )R_{\lambda}(t,y|s,x)=\\ \nonumber
         R_{\lambda}(t,y|s,x)\left(\begin{array}{cc}
          y     & 1  \\
           u+y\partial_y    &  \partial_y
         \end{array}\right )\left(\begin{array}{cc}
          t      & 1  \\
           u-\lambda+t\partial_t    &  \partial_t
         \end{array}\right )\;\;.
\end{eqnarray}
On the right hand side of the above equation, move $R_{\lambda}(ty,sy)$ from
the left side of the matrices product, to the right side. We have to change
$\partial_x \to -\partial_x$ and $x\partial_x \to -1-x\partial_x.$ Then the
equation becomes
\begin{eqnarray}
\label{mlquatum2}
\left(\begin{array}{cc}
          s      & 1  \\
           u-\lambda +s\partial_s    &  \partial_s
         \end{array}\right )\left(\begin{array}{cc}
          x      & 1  \\
           u+x\partial_x     &  \partial_x
         \end{array}\right )R_{\lambda}(t,y|s,x)=\\ \nonumber
         \left(\begin{array}{cc}
          y     & 1  \\
           u-1-y\partial_y    &  -\partial_y
         \end{array}\right )\left(\begin{array}{cc}
          t      & 1  \\
           u-\lambda-1-t\partial_t    &  -\partial_t
         \end{array}\right )R_{\lambda}(t,y|s,x)\;\;.
\end{eqnarray}
The solution is:
\begin{equation}
\label{R}
R_{\lambda}(t,y|s,x)=\rho_{\lambda}\delta(s-y)e^{y(t-x)}(t-x)^{-\lambda-1} ~.
\end{equation}

We notice that $R\sim exp(F_{\lambda})$ for $\beta=1$. Due to the Dirac
function, the solution is gauge independent, i.e. the solution does not
depend on the free variable $\xi $ from (\ref{gener}). The ${\cal
R}$-operator (\ref {intR}) becomes, after integration over y and changing
the variable $t-x=\xi$
\begin{equation}
\label{ope}
({\cal R}\psi )(s,x)
     =\rho _{\lambda}\int d\xi e^{s\xi}\,\xi^{-\lambda -1}\psi(x+\xi,s)\;,
\end{equation}
 or
\begin{equation}
\label{oper}
({\cal R}\psi )(s,x)
     =\rho _{\lambda} s^{\lambda}\int d\xi e^{\xi}\,
      \xi ^{-\lambda -1}\psi(x+s^{-1}\xi,s)\;.
\end{equation}
The branch for the many valued function $s^{\lambda}$ from (\ref{oper}) is
fixed by making a cut along $(-\infty , 0)$ and taking ${\rm arg}(s) \in
[-\pi, \pi]$. We have to specify the factor $\rho _\lambda $ and the
integration contour (in the complex $\xi$ pane) in (\ref{oper}). The
integration contour is the Hankel contour for the Gamma function
\cite{Olver}
\begin{equation}
\label{G}
\int_{-\infty}^{(0+)}e^{\xi} \xi^{-z}d\xi=\frac{2\pi i}{\Gamma(z)} ~.
\end{equation}
The previous formula inspired us to choose
\begin{equation}
\label{ro}
\rho_\lambda=\frac{1}{2\pi i}\Gamma (\lambda +1) ~.
\end{equation}
Then
\begin{equation}
\label{oper_basis}
({\cal R}_\lambda(\psi))(s,x)
      =\frac{1}{2\pi i}\Gamma(\lambda +1)s^{\lambda}\int d\xi
      e^{\xi}\xi^{-\lambda -1}\psi(x+s^{-1}\xi,s)\;.
\end{equation}

We are ready now to write the kernel of the Q-operator (\ref{K}). From
(\ref{R}) and (\ref{Qtr}) we obtain
\begin{equation}
\label{KQ}
{\cal Q}_\lambda ({\vec x}|{\vec y})=\prod_{i=1}^N w_i(\lambda ;y_{i-1},y_i,x_i)
\end{equation}
where
\begin{equation}
\label{w-i}
w_i(\lambda ;y_{i-1},y_i,x_i)
     =\frac{1}{2\pi i}\Gamma (\lambda +\beta_i )
     e^{y_i(y_{i-1}-x_i)}(y_{i-1}-x_i)^{-\lambda -\beta_i}\;.
\end{equation}

Therefore we have found an explicit form for the Baxter $Q$-operator
(\ref{KQ}). Next we are going to investigate the third property of the
$Q$-operator, namely the Baxter T-Q equation.

\section{Baxter T-Q equation}
This last paragraph aims to show that the Baxter T-Q equation is the quantum
version of the classical separation of variables (\ref{cans}). The
computation parallels the one in \cite{Sky}. Start from the left side of the
Baxter T-Q equation (\ref{TQ})
\begin{equation}
\label{TQ1}
[\tau(\lambda )Q(\lambda)\psi]({\vec x})=Tr\left[\int d{\vec t}d{\vec y}
    \left(\prod_{i=N}^1 L(\lambda|x_i,\partial_{x_i})
    R_{\lambda +\beta_i-1}(t_i,x_i|t_{i-1},y_i)\right)\psi({\vec y})\right] ~.
\end{equation}
We can integrate over $t_i$ and get
\begin{equation}
\label{Tq2}
[\tau(\lambda )Q(\lambda)\psi]({\vec x})=Tr\left[\int d{\vec y}
     \left(\prod_{i=N}^1 L(\lambda|x_i,\partial_{x_i}) w_i\right)
     \psi({\vec y})\right] ~,
\end{equation}
where $w_i$ are given by (\ref{w-i}).

Move all $w_i$ to the left using
\begin{equation}
\label{TQ5a}
L(\lambda|x_i,\partial_{x_i})w_i=w_i{\tilde L}(\lambda|x_i,\partial_{x_i})
\end{equation}
with
\begin{equation}
\label{TQ5}
{\tilde L}(\lambda|x_i,\partial_{x_i})=\left(\begin{array}{cc}
          x_i      & 1  \\
           \lambda+\beta _i+x_i\partial_{x_i}\ln w_i     &  \partial_{x_i}\ln w_i
         \end{array}\right ) ~,
\end{equation}
or
\begin{equation}
\label{TQ6}
{\tilde L}(\lambda|x_i,\partial_{x_i})=\left(\begin{array}{cc}
          x_i      & 1  \\
           -x_iy_i+\frac{(\lambda +\beta_i)y_{i-1}}{y_{i-1}-x_i}  & -y_i+\frac{(\lambda +\beta_i)}{y_{i-1}-x_i}
         \end{array}\right ) ~.
\end{equation}
At this point we can write
\begin{equation}
\label{TQ7}
[\tau(\lambda )Q(\lambda)\psi]({\vec x})
     =\int d{\vec y}\prod _{i=N}^1 w_i Tr\left(
     {\tilde L}(\lambda|x_N,\partial_{x_N}) \ldots
     {\tilde L}(\lambda|x_1,\partial_{x_1})\right)\psi({\vec y})\;.
\end{equation}
The last step is to perform a gauge transformation which leaves the trace
invariant and make the matrices ${\tilde L}(\lambda|x_i,\partial_{x_i})$
triangular, so the trace will be easy to compute
\begin{equation}
\label{TQ8}
{\tilde L}(\lambda|x_i,\partial_{x_i})
    \to N_{i}^{-1}{\tilde L}(\lambda|x_i,\partial_{x_i}) N_{i-1}\;.
\end{equation}
With the help of the following gauge matrix
\begin{equation}
\label{TQ9}
N_i=\left(\begin{array}{cc}
          1      & 0  \\
           y_i  &1
         \end{array}\right ) ~,
\end{equation}
the triangular form for ${\tilde L}(\lambda|x_i,\partial_{x_i})$ is
\begin{equation}
\label{TQ10}
N_i^{-1}{\tilde L}(\lambda|x_i,\partial_{x_i}) N_{i-1}=\left(\begin{array}{cc}
          -(y_{i-1}-x_i)      & 1  \\
          0  & \frac{\lambda +\beta_i}{y_{i-1}-x_i}
         \end{array}\right ) ~.
\end{equation}
The entries of the previous matrix can be expressed in terms of the $w_i$
(\ref{w-i})
\begin{eqnarray}
\label{TQ11}
-(y_{i-1}-x_i) \!\!\!&=&\!\!\! -(\lambda +\beta_i)\frac{w_i(\lambda-1)}{w_i(\lambda )} ~,\\
\frac{\lambda +\beta_i}{y_{i-1}-x_i} \!\!\!&=&\!\!\! \frac{w_i(\lambda+1)}{w_i(\lambda )} ~,
\end{eqnarray}
so we get for the trace
\begin{equation}
\label{TQ12}
Tr\left({\tilde L}(\lambda|x_N,\partial_{x_N}) \ldots
    {\tilde L}(\lambda|x_1,\partial_{x_1})\right)
    =\prod_{i=1}^N -(\lambda +\beta_i)\frac{w_i(\lambda-1)}{w_i(\lambda )}
    +\prod_{i=1}^N\frac{w_i(\lambda+1)}{w_i(\lambda )} ~.
\end{equation}
The last result implies the Baxter T-Q equation
\begin{equation}
\label{TQ13}
\tau (\lambda )Q(\lambda )
    =-\prod_{i=1}^N (\lambda +\beta_i) Q(\lambda -1)+Q(\lambda +1) ~.
\end{equation}
Compare (\ref{TQ13}) with the classical result (\ref{Wdres}) written in the
form
\begin{equation}
\label{Wdres1}
e^{\mu}+\tau_N-\prod_{i=1}^N(\lambda+b_i)e^{-\mu}=0 ~.
\end{equation}
The connection is obvious if we quantify the canonical pair $(\mu,\lambda )$ as
\begin{equation}
\label{quant_mu}
\mu \to \frac{d}{d\lambda} ~,\quad \lambda \to \lambda ~.
\end{equation}
Then (\ref{Wdres1}) becomes an operator acting on the Q-operator
\begin{equation}
\label{Wdres_quant}
\left(e^{\mu}+\tau_N-\prod_{i=1}^N(\lambda+b_i)e^{-\mu}\right) Q(\lambda )=0 ~,
\end{equation}
which is the T-Q equation up to a minus sign. To obtain a T-Q equation which
exactly matches the classical formula, we have to chose for $\rho _\lambda $
the one in (\ref{ro}) multiplied with $(-1)^{\lambda }$.

\section{The case N=3}

It is instructive to study the case $N=3$ which corresponds to the one-gap
potentials. The trace (\ref{monod}) of the monodromy matrix (\ref{monoL})
for $N=3$ is
\begin{equation}
\label{tau_3}
\tau_3(u)=(g_1+g_2+g_3)u + g_1 g_2 g_3 + g_1 \beta_3 + g_2 \beta_1 + g_3 \beta_2\;.
\end{equation}
The variables $g_i,\;i=1,2,3$ are (\ref{g_x}):
$g_1=X_1+x_2~,\quad g_2=X_2+x_3$ and $g_3=X_3+x_1$.
The transformation $B(\lambda )$ (\ref{Bac}, \ref{Bac-tilde}) can be explicitly found for $\lambda =-\beta_i,\;i=1,2,3$. For example, for $\lambda =-\beta_2$ we get
\begin{eqnarray}
\label{sol}
{\tilde g}_1 \!\!\!&=&\!\!\! g_3+\frac{\beta_3-\beta_2}{g_2}  ~,\\
{\tilde g}_2 \!\!\!&=&\!\!\! g_1-\frac{\beta_3-\beta_2}{g_2}
       +\frac{\beta_1-\beta_2}{g_3+\frac{\beta_3-\beta_2}{g_2}} ~,\\
{\tilde g}_3 \!\!\!&=&\!\!\! g_2-\frac{\beta_1-\beta_2}{g_3+\frac{\beta_3-\beta_2}{g_2}} ~.
\end{eqnarray}

This transformation can be recovered from the B\"acklund transformations
$T_k,\;k=1,2,3$ from \cite{Veselov,Adler}. Recall that $T_k$ is given by
\begin{eqnarray}
\label{Ba_ves}
T_k(g_{k\pm 1}) \!\!\!&=&\!\!\! g_{k\pm 1}\pm \frac{\beta_{k+1}-\beta_k}{g_k} ~,\\
T_k(\beta _k) \!\!\!&=&\!\!\! \beta_{k+1} ~,\\
T_k(\beta_{k+1}) \!\!\!&=&\!\!\! \beta_k ~,
\end{eqnarray}
the remaining $\beta _j$ and $g_j$ being not changed. We also need to
introduce the shift $S$ acting as
\begin{eqnarray}
\label{S}
S(\beta_i) \!\!\!&=&\!\!\! \beta_{i-1} ~,\\
S(g_i) \!\!\!&=&\!\!\! g_{i-1}~.
\end{eqnarray}
In terms of these last transformations, we can write
\begin{equation}
\label{T}
B(-\beta_2)=T_2 S T_1 ~.
\end{equation}
We cannot recover $T_k$ from $B(\lambda )$ because of the difference in
nature between these transformations. $T_k$ transforms the parameters
$\beta_j$ so it changes solutions of one system of equations (\ref{chain})
to solutions of another system of the same type (\ref{chain}). The
transformations $B(\lambda)$ change the solutions of the same system
(\ref{chain}) among themselves. In this respect $B(\lambda )$ is an
auto-B\"acklund transformation.

We can try to solve the discrete Riccati equation (\ref{linear}) for
$\psi_i$. In this case, we will get $\psi _1,\cdots, \psi_3$ in terms of
$\psi_0$ and $\psi_4 $
\begin{eqnarray}
\label{psi_3}
\psi_1 \!\!\!&=&\!\!\! \frac{1}{Z}\left[\psi_4 -\psi_0 (g_2g_3+\lambda+\beta_3)(\lambda+\beta_1)\right] ~,\\
\psi_2 \!\!\!&=&\!\!\! \frac{1}{Z}\left[\psi_4 g_1+\psi_0g_3(\lambda+\beta_2)(\lambda +\beta_1)\right] ~,\\
\psi_3 \!\!\!&=&\!\!\! \frac{1}{Z}\left[\psi_4(g_1g_2+\lambda+\beta_2)
   -\psi_0(\lambda+\beta_1)(\lambda+\beta_2)(\lambda+\beta_3)\right] ~,
\end{eqnarray}
with $Z=g_1g_2g_3+g_1(\lambda+\beta_3)+g_3(\lambda+\beta_2)$. It obvious that
$z_i=\psi_{i+1}/\psi_i$ will depend on $\psi _4$ and $\psi_0$ only through their
ratio $\psi_4/\psi_0.$ This ratio is not a free parameter because the periodic
boundary condition $z_0=z_3$ imposes a restriction on it.

Though the $\tau$ -functions are not one of the major players of this paper,
it is worthwhile to mention the connection it has with canonical transformations.

The $\tau$ -functions for the dressing chain were reported in \cite{Okamoto} for
$N=3$.  For $N=3$ case, the dressing chain (\ref{chain}) written in variables $g_i$,
(\ref{g}), is:
\begin{eqnarray}
\label{g3}
g_0' \!\!\!&=&\!\!\! -g_0 (g_1-g_2) + \beta_0 - \beta_1~, \nonumber \\
g_1' \!\!\!&=&\!\!\! -g_1 (g_2-g_0) + \beta_1 - \beta_2~,  \\
g_2' \!\!\!&=&\!\!\! -g_2 (g_0-g_1) + \beta_2 - \beta_0~. \nonumber
\end{eqnarray}
With the change of variables
\begin{eqnarray}
\label{g-to-F}
g_0 \!\!\!&=&\!\!\! F_1' -F_2' +c ~,\nonumber \\
g_1 \!\!\!&=&\!\!\! F_2' -F_0' +c ~, \\
g_2 \!\!\!&=&\!\!\! F_0' -F_1' +c ~,\nonumber
\end{eqnarray}
where the constant $c$ is given by $3 \, c = g_0+g_1+g_2$, the system of equations
(\ref{g3}) transforms into
\begin{eqnarray}
\label{F}
F_0'' + F_1'' + (F_0' - F_1')^2 -c\,(F_0' - F_1') + \beta_1 \!\!\!&=&\!\!\! 0~, \nonumber \\
F_1'' + F_2'' + (F_1' - F_2')^2 -c\,(F_1' - F_2') + \beta_2 \!\!\!&=&\!\!\! 0~, \\
F_2'' + F_0'' + (F_2' - F_0')^2 -c\,(F_2' - F_0') + \beta_0 \!\!\!&=&\!\!\! 0~.\nonumber
\end{eqnarray}
The $\tau$ --functions $\tau_0, \tau_1, \tau_2$ are now given by
\begin{equation}
\label{F-tau}
F_0 = \log \tau_0~,~~F_1 = \log \tau_1~,~~F_2 = \log \tau_2~~.
\end{equation}
In terms of the $\tau$ -functions, the dressing chain becomes a Hirota type system
of equations:
\begin{eqnarray}
\label{tau}
(D_x^2 -c D_x +\beta_1)\, \tau_0 \cdot \tau_1 \!\!\!&=&\!\!\! 0~, \nonumber \\
(D_x^2 -c D_x +\beta_2)\, \tau_1 \cdot \tau_2 \!\!\!&=&\!\!\! 0~, \\
(D_x^2 -c D_x +\beta_0)\, \tau_2 \cdot \tau_0 \!\!\!&=&\!\!\! 0~, \nonumber
\end{eqnarray}
where for a polynomial $P$, the operator $P(D_x)$ is defined as
\begin{equation}
\label{P}
P(D_x)\,F(x)\cdot G(x) = P(\partial_y)\,F(x+y)G(x-y)|_{y=0}~.
\end{equation}

At this point the goal is to work the canonical transformation (\ref{Bac}) and
(\ref{Bac-tilde}) in terms of the $\tau$ -functions.
The canonical transformation (\ref{Bac}), (\ref{Bac-tilde}) namely
\begin{equation}
\label{g1}
g_i = z_i - \frac{\lambda + \beta_i}{z_{i-1}}
\end{equation}
and
\begin{equation}
\label{g2} \tilde{g}_i = g_{i+1} + z_{i-1} -z_{i+1}~,
\end{equation}
cannot be written explicitly as a formula which comprise only $\tilde{g}_i$ and
$g_i$. Therefore we use the following strategy: given $g_i$, we have to solve
for $z_i$ in (\ref{g1}), and then obtain the transformed variables $\tilde{g}_i$
with the aid of (\ref{g2}). As a result, the canonical transformation for the
$\tau$ -functions will be written in terms of the variables $z_i$.

First, by simple manipulations of (\ref{g3}) and (\ref{g-to-F}) we get
\begin{equation}
\label{F0} F_0'' = g_1 g_2 - c^2 - \frac{\beta_0 + \beta_1 - \beta_2}{2}~,
\end{equation}
and similarly for $F_1''$ and $F_2''$. Using (\ref{F-tau}) we get
\begin{equation}
\label{log-tau}
\left( \log e^{-\gamma_0 x^2} \tau_0 \right)'' = g_1 g_2~,
\end{equation}
and similarly for $\tau_1$ and $\tau_2$. Here
$\gamma_0 = -\frac{1}{2} \left(c^2 + \frac{\beta_0 +\beta_1 -\beta_2}{3}\right)$~.

After we apply the canonical transformation we obtain the function $\tilde{\tau_0}$
given by
\begin{equation}
\label{log-tau-tilde}
\left( \log e^{-\gamma_0 x^2} \tilde{\tau_0} \right)'' =
\tilde{g_1} \tilde{g_2} ~.
\end{equation}
Now it is easy to write the canonical transformation for $\tau_0, \tau_1, \tau_2$ in
terms of the variable $z_i$. Use (\ref{g1}) and (\ref{g2}) in (\ref{log-tau}) and
(\ref{log-tau-tilde}) and get:
\begin{eqnarray}
\label{tau1}
\left( \log e^{-\gamma_0 x^2} \tau_0 \right)'' \!\!\!&=&\!\!\!
\left(z_1-\frac{\lambda+\beta_1}{z_0} \right)\left(z_2-\frac{\lambda+\beta_2}{z_1}\right)~,
\nonumber \\
\left( \log e^{-\gamma_1 x^2} \tau_1 \right)'' \!\!\!&=&\!\!\!
\left(z_2-\frac{\lambda+\beta_2}{z_1} \right)\left(z_0-\frac{\lambda+\beta_0}{z_2}\right)~,\\
\left( \log e^{-\gamma_2 x^2} \tau_2 \right)'' \!\!\!&=&\!\!\!
\left(z_0-\frac{\lambda+\beta_0}{z_2} \right)\left(z_1-\frac{\lambda+\beta_1}{z_0}\right)~,
\nonumber
\end{eqnarray}
and
\begin{eqnarray}
\label{tau2}
\left( \log e^{-\gamma_0 x^2} \tilde{\tau_0} \right)'' \!\!\!&=&\!\!\!
\left(z_0-\frac{\lambda+\beta_2}{z_1} \right)\left(z_1-\frac{\lambda+\beta_0}{z_2}\right)~,
\nonumber \\
\left( \log e^{-\gamma_1 x^2} \tilde{\tau_1} \right)'' \!\!\!&=&\!\!\!
\left(z_1-\frac{\lambda+\beta_0}{z_2} \right)\left(z_0-\frac{\lambda+\beta_1}{z_0}\right)~, \\
\left( \log e^{-\gamma_2 x^2} \tilde{\tau_2} \right)'' \!\!\!&=&\!\!\!
\left(z_2-\frac{\lambda+\beta_1}{z_0} \right)\left(z_0-\frac{\lambda+\beta_2}{z_1}\right)~,
\nonumber
\end{eqnarray}
with
$$
\gamma_0 = -\frac{1}{2} \left(c^2 + \frac{\beta_0 +\beta_1 -\beta_2}{3}\right)~,~
\gamma_1 = -\frac{1}{2} \left(c^2 + \frac{\beta_1 +\beta_2 -\beta_0}{3}\right)~,~
\gamma_2 = -\frac{1}{2} \left(c^2 + \frac{\beta_2 +\beta_0 -\beta_1}{3}\right)~~.
$$
The meaning of the transformations as presented in (\ref{tau1}) and (\ref{tau2}) is that
first we must factorize like in (\ref{tau1}) and then use the factorization variables
$z_i$ to obtain the transformed $\tau$ -functions, like in (\ref{tau2}).

For the case $N=3$, the canonical transformation must be carried on three times, each
time with another $\lambda$ in order to obtain separated canonical variables,
see (\ref{cans}).
In variables $g_i$, these transformations read as
\begin{eqnarray}
\label{g-lambda}
\tilde{g}_i \!\!\!&=&\!\!\! g_{i+1} +z_{i-1} - z_{i+1}~~,~~
    g_i = z_i - \frac{\lambda_1+\beta_i}{z_{i-1}}~, \nonumber \\
\tilde{\tilde{g}}_i\!\!\!&=&\!\!\! \tilde{g}_{i+1} +\tilde{z}_{i-1} - \tilde{z}_{i+1}~~,~~
    \tilde{g}_i = \tilde{z}_i - \frac{\lambda_2+\beta_i}{\tilde{z}_{i-1}} ~, \\
\tilde{\tilde{\tilde{g}}}_i\!\!\!&=&\!\!\! \tilde{\tilde{g}}_{i+1} +\tilde{\tilde{z}}_{i-1} - \tilde{\tilde{z}}_{i+1}~~,~~
    \tilde{\tilde{g}}_i = \tilde{\tilde{z}}_i - \frac{\lambda_3+\beta_i}{\tilde{\tilde{z}}_{i-1}} ~. \nonumber
\end{eqnarray}
If we add $\mu_1, \mu_2, \mu_3$ given by (\ref{zpr}),
\begin{eqnarray}
\label{g-mu}
e^{\mu_1}\!\!\!&=&\!\!\!-z_1 z_2 z_3 ~,\nonumber \\
e^{\mu_1}\!\!\!&=&\!\!\!-\tilde{z}_1 \tilde{z}_2 \tilde{z}_3~, \\
e^{\mu_1}\!\!\!&=&\!\!\!-\tilde{\tilde{z}}_1 \tilde{\tilde{z}}_2 \tilde{\tilde{z}}_3 ~,\nonumber
\end{eqnarray}
the variables $(\lambda_1 ,\lambda_2 ,\lambda_3 , \mu_1 , \mu_2 , \mu_3)$ are separated
canonical variables. The time evolution of these variables, inherited from the dressing
chain (\ref{g3}) is such that, (\ref{Wdres}),
\begin{equation}
\label{W}
e^{2\mu_1} +  e^{\mu_1}-(\lambda_1+\beta_1)(\lambda_1+\beta_2)(\lambda_1+\beta_3)= 0~,
\end{equation}
and similarly for the pairs $(\lambda_2,\mu_2)$ and $(\lambda_3,\mu_3)$.
Here $\tau_N(\lambda_1)$ is given by (\ref{tau_3}) with $\lambda_1$ instead of $u$.

If we use the notation
$
\tau^{(1)} = \tau~,~~ \tau^{(2)} = \tilde{\tau}~,~~\tau^{(3)} = \tilde{\tilde{\tau}}
$
and similarly for $z_i$,
the transformations (\ref{g-lambda}) for the $\tau$ -functions, can be compactly
expressed as follows:

factorization:
    $$
    \left( \log e^{-\gamma_i x^2} \tau_i^{(k)} \right) ''
    = \left( z_{i+1}^{(k)} - \frac{\lambda_k + \beta_{i+1}}{z_i^{(k)}} \right)
      \left( z_{i+2}^{(k)} - \frac{\lambda_k + \beta_{i+2}}{z_{i+1}^{(k)}} \right)~,
    $$

transformation:
    $$
    \left( \log e^{-\gamma_i x^2} \tau_i^{(k+1)} \right) ''
    = \left( z_{i}^{(k)} - \frac{\lambda_k + \beta_{i+2}}{z_{i+1}^{(k)}} \right)
      \left( z_{i+1}^{(k)} - \frac{\lambda_k + \beta_i}{z_{i+2}^{(k)}} \right)~.
    $$
Here $i=0,1,2$ modulo $3$, and $k=1,2,3$.
The variables $\mu$ are given by (\ref{g-mu}) as before.

\section{Hamiltonian flow for the case $N=3$}
Here we discuss the time dependence of the canonical variables $(X_i,x_i).$
We regard the variable $x$ (i.e. the {\it space} variable) in the system of
equations (\ref{chain}) as being a time $t $ for the Hamiltonian flow. The
Hamiltonian that governs the motion in time is in variables $f_i$ (\ref{g})
given by
\begin{equation}
\label{Ham_3}
H=\frac{1}{3}(f_1^3+f_2^3+f_3^3+\beta_1f_1+\beta_2f_2+\beta_3f_3)\;.
\end{equation}
It is useful to list all the variables which appeared so far
\begin{eqnarray}
\label{g_f}
g_1 \!\!\!&=&\!\!\! f_1+f_2\;,\;f_1=\frac{1}{2}(g_1-g_2+g_3)\;,\;g_1=X_1+x_2 ~,\\
g_2 \!\!\!&=&\!\!\! f_2+f_3\;,\;f_2=\frac{1}{2}(g_2-g_3+g_1)\;,\;g_2=X_2+x_3 ~,\\
g_3 \!\!\!&=&\!\!\! f_3+f_1\;,\;f_3=\frac{1}{2}(g_3-g_1+g_2)\;,\;g_3=X_3+x_1 ~.
\end{eqnarray}
The evolution of the variable $x_1$ in time is given by
\begin{equation}
\label{ev_x}
\frac{dx_1}{dt}=\{x_1,H\}\;.
\end{equation}
The Poisson bracket can be computed for two arbitrary functions $f$ and $g$
from
\begin{equation}
\label{Pois}
\{f,g\}=\sum_{\alpha ,\beta}\frac{\partial f}{\partial y_\alpha }
   \frac{\partial g}{\partial y_\beta }\{y_{\alpha},y_{\beta }\} ~,
\end{equation}
where $(y_1,\ldots,y_{2N})=(X_1,\ldots ,X_N,x_1,\ldots,x_N)$. In this way we
arrive at
\begin{equation}
\label{Poi_x}
2\frac{dx_1}{dt}=(f_1^2+f_2^2-f_3^2+\beta _1+\beta _2-\beta _3)\{x_1,X_1\} ~,
\end{equation}
which gives the evolution for $x_1$ knowing that the variables $(X_1,x_1)$
are canonical, i.e.
\begin{equation}
\label{ca}
\{x_1,X_1\}=1 ~.
\end{equation}
{}From the paper of Veselov and Shabat \cite{Veselov} we even know the
solutions for the dressing chain in terms of the elliptic Weierstrass ${\cal
P}$-functions:
\begin{equation}
\label{Wei}
f_i(t)=\frac{1}{2} \frac{{\cal P}^{'}(t+a_i)
   -{\cal P}^{'}(b_i)}{{\cal P}(t+a_i)-{\cal P}(b_i)} ~.
\end{equation}
We can integrate (\ref{Poi_x}) to
\begin{eqnarray}
\label{x}\nonumber
2x_1(t) \!\!\!&=&\!\!\! \xi(t+a_3+b_3)+\xi (t+a_3)+\xi (b_3)-\xi (t+a_1+b_1)
  -\xi (t+a_1)-\xi (b_1)-\\
& &\xi (t+a_2+b_2)-\xi _(t+a_2)-\xi (b_2)+\beta_1+\beta_2-\beta_3 ~.
\end{eqnarray}
Here $\xi' = - {\cal P}, b_i={\cal P}(\beta _i)$ and $a_{i+1}-a_i=b_i\;$.

In the quantum case the Hamiltonians are generated from the Lax matrix
(\ref{quntL}). The trace of the monodromy matrix is:
\begin{equation}
\label{ta_quant}
\tau_3(u)=H_1u+H_3
\end{equation}
where
\begin{eqnarray}
\label{H_quant}
H_1&=&x_1+x_2+x_3+\partial _{x_1}+\partial _{x_2}+\partial _{x_3} ~,\\
H_3&=&\partial _{x_1}\partial _{x_2}\partial _{x_3}
 +x_1\,\partial _{x_1}\partial _{x_3}+x_2\,\partial _{x_2}\partial _{x_1}
 +x_3\,\partial _{x_3}\partial _{x_2} \nonumber\\
{~} &+&\!\!\! x_2x_1\,\partial _{x_1}+x_3x_2\,\partial _{x_2}+x_1x_3\,\partial _{x_3}
  +x_1x_2x_3 ~~.
\end{eqnarray}

An interesting feature of the quantum case is the absence of evolution of
the variables $g_i=\partial _{x_i} +x_{i+1}$ due to
\begin{equation}
\label{no_ev}
[g_i,H_3]=0 ~.
\end{equation}
This means that there is no quantum analog of the dressing chain in terms of
the variables $g_i$. The only surviving variables are $X_i=\partial _{x_i}$
and $x_i$ which each separately evolve in time under the Hamiltonian $H_3$.

\section{Conclusions}

For the finite-gap potentials we have shown that the Darboux transformations
can be viewed as canonical transformations if we apply the method of
separation of variables proposed by Sklyanin. Not only the finite-gap case
(N=odd, $\alpha =0$ ) is interesting but also the other cases. For example
the spectrum of the potentials which are solution of the chain in the case
 $\alpha \ne 0$ and arbitrary $N$ is described as following.
The ground state is at zero energy; the next $(p-1)$ eigenvalues
are $E_l = \sum_{k=0}^{l} \alpha_k~,~l=0,1,\ldots,(p-2)~ $, and all other
eigenvalues are obtained by adding arbitrary multiples
of the quantity $\alpha \equiv \alpha_0+\alpha_1+\cdots+\alpha_{p-1}$. The general formula for the excited energy levels is: \cite{Sukhatme}
\begin{equation}
n \alpha + \sum_{k=0}^l \alpha_k ~~;
~\{n=0,1,2,\ldots,\infty~~;~~l=0,1,\ldots,(p-1)\}~~.
\label{ev}
\end{equation}
The above potentials (also called {\em cyclic shape invariant potentials})
are a direct generalization of the harmonic oscillator.
For $N=3$ the potentials are Painlev\'e transcendents \cite{Veselov}.

In this work we have also shown that there exists a quantum version of the dressing
chain, namely the time evolution of the variables $(X,x)$ under the Hamiltonian
$H_1$. From here there are many ways to proceed. One way is along Bethe-Ansatz
procedure. It will be interesting to find the spectrum of the quantum Hamiltonians.
Also, each Hamiltonian $H_i$ has its own time $t_i$ for evolution, so there must be
a $\tau $-function which depends on all time variables $\tau (t_1,t_2,...,t_N)$. The
$\tau $-function for the dressing chain was reported in \cite{Okamoto} for $N=3$
in connection with the Painlev\'{e} equations. See also \cite{Yamada}.

We have analyzed the role of $\tau$ -functions in connection with the factorization
method and canonical transformation. The $\tau$ -functions are an important tool in
understanding integrable systems and we believe that the connections between the
$\tau$ -functions and the Darboux transformations is worth to be studied further.

Finally, a word about KdV. The finite-gap potential
theory provides solutions of the periodic boundary problems for KdV
equation. The KdV equation
\begin{equation}
u_{xt}=u_{xxx}-6uu_{xx}
\end{equation}
is a partial differential equation in two variables. One variable $x$ we
interpreted as a time variable associated with the Hamiltonian $H_1$. To
what Hamiltonian is the second variable, i.e. t, associated? For $\beta_i=0$
the Hamiltonian is $H_5^2-H_1H_3$. This result is buried in the paper
\cite{Weiss}. What is the meaning of the Sklyanin separation of variables
for KdV equation, both classical and quantum? From the conformal field point
of view, the quantum KdV and the T-Q relation was already studied in the
paper of \cite{Zam}.

In conclusion, the Darboux transformation together with the new approach of
the separation of variables is a promising research direction.

\section*{Acknowledgments}
\noindent
O.L. is grateful to P.B. Wiegmann for useful discussions.

\end{document}